# Observation of a Snap-Through Instability in Graphene

Scott Scharfenberg, [1] Nikhita Mansukhani, [1] Cesar Chialvo, [1] Richard L. Weaver, [2] and Nadya Mason[1]

[1] *Department of Physics and Materials Research Laboratory, University of Illinois at Urbana-Champaign, 1110 West Green Street, Urbana, Illinois 61801, USA*

[2] *Institute for Condensed Matter Theory, University of Illinois at Urbana-Champaign, 1110 West Green Street, Urbana, Illinois 61801, USA*

## Abstract

We examine the competition between adhesive and bending energies for few-layered graphene samples placed on rigid, microscale-corrugated substrates. Using atomic force microscopy, we show that the graphene undergoes a sharp "snap-through" transition as a function of layer thickness, where the material transitions between conforming to the substrate and lying flat on top of the substrate. By utilizing the critical snap-through thickness in an elasticity model for the FLG's bending, we extract a value for graphene-surface adhesion energy that is larger than expected for van der Waals forces.

Graphene's exceptional properties give it great potential for applications in electronic and mechanical devices, and in composite materials[1,2]. Because graphene is thin and flexible, its interface with other materials can determine its extrinsic morphology[3-5] which in turn influences its electronic and mechanical properties[6,7]. Understanding graphene's morphology at the interface with other materials is therefore crucial in tailoring graphene's properties for device and materials applications. Of particular interest is a sudden morphological change – a "snap-through transition" – that has been predicted to occur when graphene adheres to a rough or corrugated surface and experiences strain[8,9]. At the snap-through point, graphene's surface adhesion energy equals its conformed strain energy, and the material transitions between conforming to the substrate and lying on top of the substrate, as depicted in Fig. 1. The existence of a snap-through transition is relevant for characterizing the adhesion and mechanical properties of graphene-material interfaces, and could also form the basis of novel substrate-regulated graphene devices[9]. Although previous measurements of graphene exfoliated onto corrugated elastic substrates[10] were consistent with a snap-through transition, a clear transition in graphene has not yet been measured.

In this Letter, we demonstrate that a clear snap-through transition occurs for few-layer graphene (FLG) exfoliated onto micro-scale corrugated metallic surfaces. The FLG's strain energy is tuned via its thickness, as its bending rigidity increases with the number of layers[11]. The topographies of various thicknesses of FLG on the corrugated surface are determined using an atomic force microscope (AFM). We find that the FLG changes from closely conforming to the substrate to remaining nearly flat on top of the substrate—i.e., a snap-through transition occurs—when the FLG thickness is greater than a critical number of layers. The snap-through transition is remarkably sharp, and occurs at a surprisingly large thickness of 61 layers. Using an elasticity model for the FLG's bending, we extract a graphene-surface adhesion energy of ~ 1.1 eV/nm$^2$. This value is larger than typical for micro-mechanical systems[12], and is consistent with recent studies indicating that graphene's adhesion can be larger than typical for solid-solid van der Waals forces[13].

To create the samples, we start with the top, grooved Al layer of a commercial compact disc, which has regular rounded grooves that are 200 nm deep and have a peak-to-peak separation of 1.5 μm. The thin Al from the compact disc is glued to a thicker metal disc, for mechanical stability, then sputter coated with a 40 nm Pd/Au mixture. The Pd/Au coating creates a relatively clean substrate surface. FLG is then mechanically exfoliated onto the grooved Pd/Au. Following exfoliation of FLG, candidate samples are first located using optical microscopy, then imaged on an Asylum Research Cypher AFM. Figure 1(c) displays a topographic image of FLG on one of the grooved substrates, for an FLG flake that closely conforms to the corrugations.

Figure 2 shows typical AFM height data for different thicknesses of FLG on the corrugated Pd/Au substrate. FLG flakes of various thicknesses are randomly produced by the exfoliation process; the thickness of each flake was measured by AFM height scans along the top of a groove. Figure 2(a) shows measurements of a 46-layer thick piece of FLG that fully conforms to the grooves of the substrate. (The rigid substrate does not deform, in contrast to the case of FLG on corrugated PDMS[10]). The lower part of the figure shows the height across multiple grooves on the substrate alone (blue line) and across the FLG that is over the substrate (red line). It is clear that the centers and magnitudes of the heights for the FLG and the substrate are nearly identical, implying that the FLG conforms to the corrugations. In contrast, Fig. 2(c) shows an 88-layer piece of FLG that does not conform to the corrugations. In this case, the height data shows that while the substrate alone is still strongly corrugated (right part of blue line), the FLG on top of the substrate is not corrugated (right part of red line). It is also clear from the average height of the FLG (red line) that the position of the FLG is offset to the top of the substrate corrugations. The data thus show that 88-layer FLG does not conform to the substrate, but rather lies on top of it. Similar AFM height data was taken for fifteen pieces of FLG, on two different chips, for FLG thicknesses between 10 and 90 layers. Most FLG flakes either conformed or lay flat; however, flakes having thicknesses between 60 and 62 layers were found to partially conform to the substrate, as can be seen in Fig. 2(b).

The data for all samples can be depicted by following the nomenclature of Ref[9] and introducing a dimensionless adhesion factor $A = A_g/A_s$, where $A_g$ and $A_s$ are the peak-to-peak corrugation heights of the FLG and the substrate, respectively. Figure 3 shows $A$ as a function of FLG thickness for all measured samples. It can be seen that a steep transition between fully conformal ($A \sim 1$) to fully flat ($A \sim 0$) behavior occurs around a critical thickness of $n_c = 61$ layers; this is the snap-through transition. The transition width is narrow, occurring between 60 layers and 62 layers. The discontinuity in the comformability seems to depend only on the FLG thickness, and not on the specific details of sample size, shape, or position. Our experimental observations are consistent with predictions of Li *et al* in [8] of a discontinuous snap-through transition that occurs when FLG's elastic energy of bending (which increases with its thickness) exceeds its adhesion energy.

Our experimental observations can be used to deduce the energy of adhesion between the FLG and the metal substrate. It is difficult to quantitatively compare our results to predictions of Refs [8,9], as the authors consider a regime in which the forces are assumed to be van der Waals, and in which corrugation amplitudes are smaller than, or comparable to, the FLG thickness. Instead we introduce a (simpler) theory here, and start by recognizing that the adhesion energy of a fully conforming piece of FLG can be no less than the elastic energy of FLG bending. An FLG sheet bent into a sinusoidal corrugation with out of plane displacement y = H cos(kx), so as to conform to the corrugations of a substrate, has an areal bending energy density given by

$$E \; = \; (1/4) \, D \, (y''(x))^2 = (1/4) \, DH^2 \, k^4 \cos^2(kx) \qquad\qquad (1)$$

which averages to

$$\bar{E} \; = \; (1/4) \, D \, H^2 \, k^4 \qquad .$$

In (1), $D$ is the bending rigidity of the FLG sheet, which can be modeled for an isotropic continuum[11] as $D = Yh^3/(12(1-\nu^2))$ [14], where $h$ is the FLG thickness, $Y$ is its in-plane Young's modulus, and $\nu$ is its Poisson ratio.[i] It is helpful to write $D = dn^3$ where n is the number of layers and $d$ is a scaled bending rigidity. By taking $Y = 900$ GPa[15] and an average layer thickness of 0.335 nm, the scaled rigidity $d$ is estimated to be about 18 eV, nearly independent of Poisson ratio. At the observed critical value of $n = 61$, and also using $\nu = 0.16$, $H = 60$ nm and $k = 4.2$/micron, we evaluate $\bar{E}$ to be at least 1.1 eV/nm$^2$. This is then an estimate of the adhesion energy between the FLG and the metal substrate.   This value is larger than some predictions for van der Waals forces (e.g., 0.04 eV/nm$^2$ for FLG on SiO$_2$[16]), though consistent with the FLG-Si adhesion energy of 0.9 eV/nm$^2$ obtained by measuring nanoscale FLG blisters[17]. Recent measurements of FLG-SiO$_2$ adhesion using pressurized blister tests indicate adhesion energies of ~ 1.9 ev/nm$^2$ for two to five-layer graphene on SiO$_2$[13]. In the latter experiment, the strong adhesion was ascribed to the significant conformation of graphene to substrates, leading to liquid-like, rather than solid-like, surface interactions. We note that we obtain large values of FLG adhesion even for much thicker FLG than have been measured previously. For our measurements, it is also possible that metallic bonding is relevant.

---

[i] Eq. (1) requires modification if kH is comparable to or greater than unity. In our case $kH$ is 0.25 and we neglect such considerations; a quantitative estimate suggests that this neglect leads to an overestimate of the average bending energy by about 10%. It also requires modification if the periodic corrugations are non-sinusoidal. As bending energy scales with $k^4$, higher harmonics can be significant. A quantitative estimate for the effect, based on an measurements of the Fourier components of the Au corrugations (amplitudes at $k$, $2k$ and $3k$ being found in a ratio of 8.4:1:1 and higher harmonics being negligible), suggests that our neglect of the higher harmonics leads to (1) being an underestimate of the bending energy by a factor of more than two. There is some uncertainty here as the FLG may have debonded at the points of highest curvature. A purely sinusoidal substrate would permit more quantitatively precise estimates of adhesion energy. Further modifications would be required if the layers of the FLG slipped relative to each other. The lack of slip is justified a posteriori by an estimate of the shear strain (at the midplane of a fully conformed n=60 FLG) of $(nd)2k3H/4$=4 x10$^{-4}$, well within the elastic limit of defectless solids.

In conclusion, we observe that graphene undergoes a sharp transition between conforming to the substrate and lying on top of the substrate at a critical layer thickness/bending energy. These observations are consistent with predictions, and allow the graphene-substrate adhesion energy to be extracted. The extracted value of the adhesion energy is large, and consistent with measurements of strong adhesion using blister tests. We note that our method of extracting adhesion energies is less complicated, in both implementation and analysis, than blister tests. In general, our results introduce a straightforward method of determining adhesive properties in small materials. The methods described here can be applied to various membranes placed on surfaces having different corrugation amplitudes and shapes, and composed of different materials.

We thank Scott Maclaren (UIUC MRL/CMM). This work was supported by NSF Grant No. DMR-0644674, and was carried out in part in the Frederick Seitz Materials Research Laboratory Central Facilities, University of Illinois.

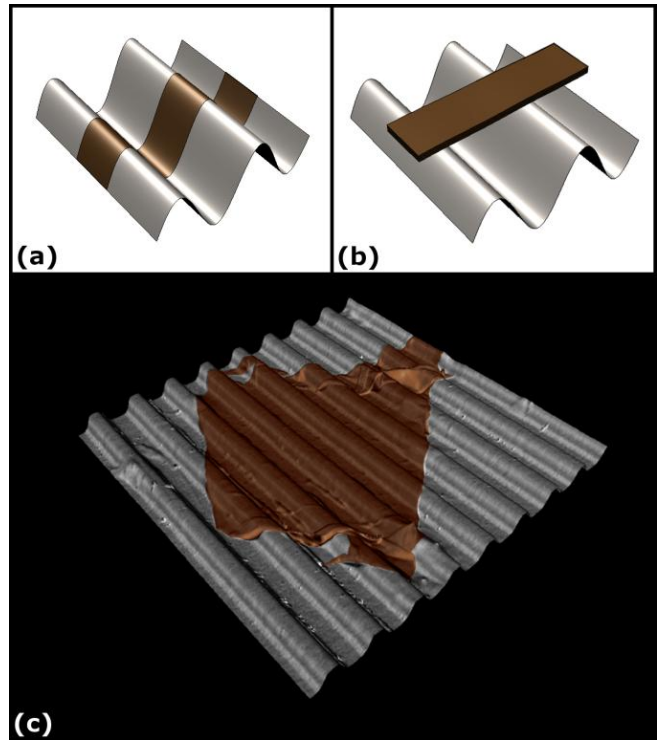

Figure 1: (a) Schematic representation of FLG having thickness below the critical number of layers, $n_c$, for the snap-through transition. Here, adhesive energy is larger than bending energy. (b) Schematic representation of FLG having thickness above $n_c$, where bending energy is larger than adhesive energy. (c) AFM topographic image (height plus superimposed phase) of 24-layer graphene that closely conforms to the corrugated substrate.

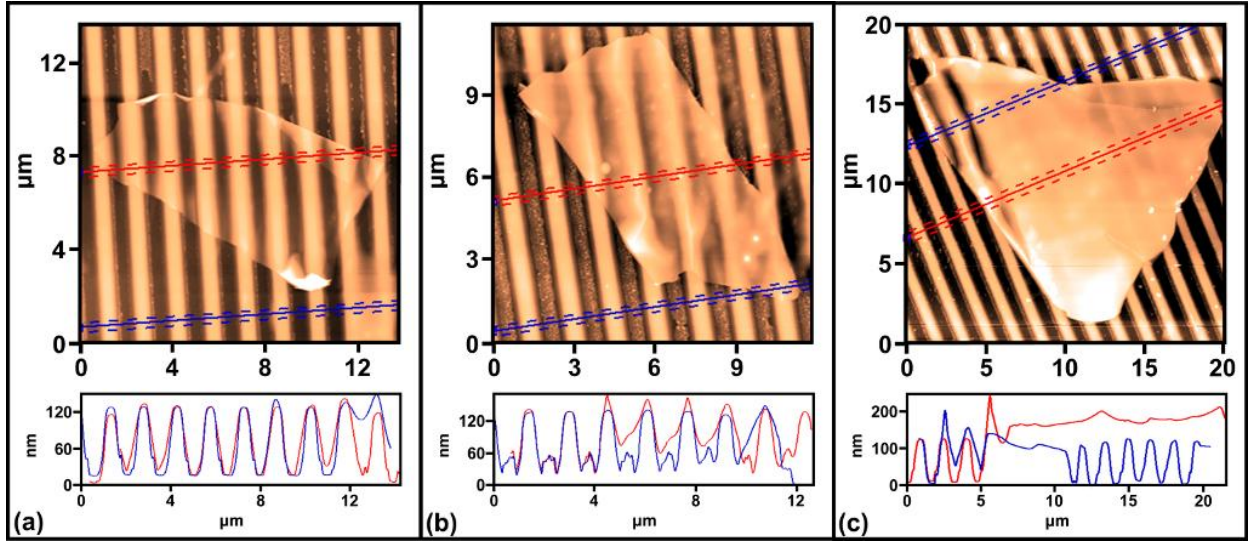

Figure 2: AFM height data showing 2D topography (upper images) and line scans (lower curves). Red and blue line scans give heights averaged between the dashed lines shown in the topographic images. (a) 46-layer graphene, which conforms to substrate corrugations. (b) 60-layer graphene, which partially adheres to the substrate. This thickness falls along the snap-through transition line. (c) 88-layer graphene, which sits on top of substrate.

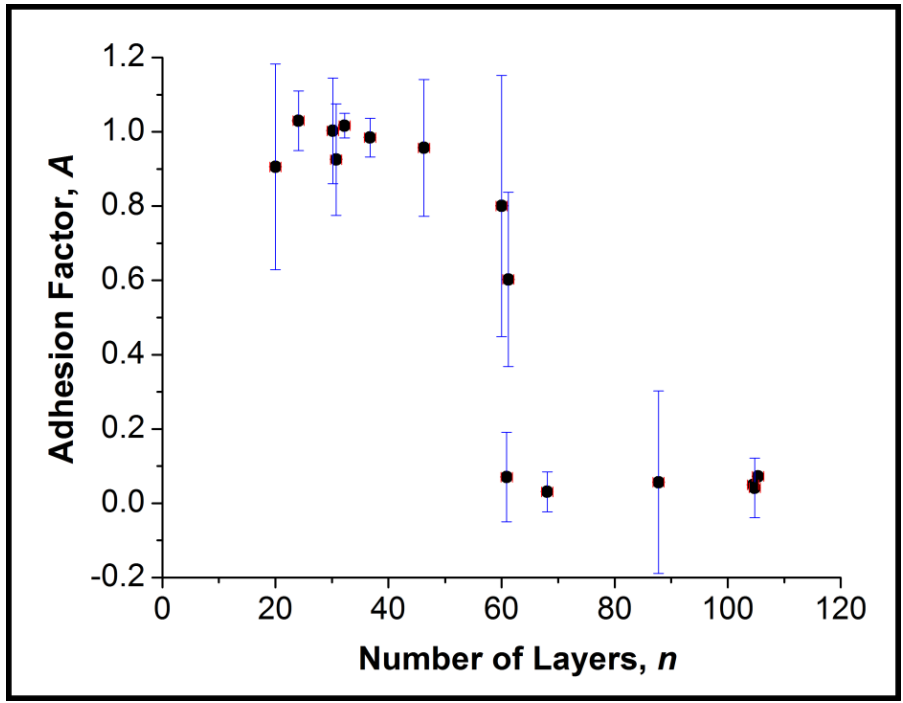

Figure 3: Adhesion factor, $A = A_g/A_s$, as a function of FLG thickness. $A_g$ and $A_s$ are peak-to-peak amplitudes of the graphene and substrate, respectively. Here, $A = 1$ corresponds to FLG that fully conforms to the substrate, while $A = 0$ corresponds to FLG that is flat. Note the sharp transition between the states for a critical thickness of $n_c \approx 61$ layers.